\documentclass[aps, pra, twocolumn, lengthcheck]{revtex4-1}
\usepackage{amsfonts, amssymb, amsmath, graphicx, comment, bm, slashed, caption, subcaption, dcolumn}

\newcommand{\beq}{\begin{eqnarray}}
\newcommand{\eeq}{\end{eqnarray}}
\newcommand{\beqa}{\begin{eqnarray}}
\newcommand{\eeqa}{\end{eqnarray}}
\newcommand{\Lc}{{\cal{L}}}
\newcommand{\Uc}{{\cal{U}}}

\def\braket#1{\mathinner{\langle{#1}\rangle}}

\begin{document}

\title{Asymmetric pairing of realistic mass quarks and color neutrality in the \\ Polyakov--Nambu--Jona-Lasinio model of QCD}

\author{Philip D. Powell}
\author{Gordon Baym}

\affiliation{Department of Physics, University of Illinois at Urbana-Champaign, 1110 West Green Street, Urbana, Illinois 61801, USA}

\date{\today}

\begin{abstract}
We investigate the effects of realistic quark masses and local color neutrality on quark pairing in the three-flavor Polyakov--Nambu--Jona-Lasinio model.  While prior studies have indicated the presence of light flavor quark (2SC) or symmetric color-flavor-locked (CFL) pairing at low temperatures, we find that in the absence of a local color neutrality constraint the inclusion of the Polyakov loop gives rise to phases in which all quark colors and flavors pair, but with unequal magnitudes.  We study this asymmetric color-flavor-locked (ACFL) phase, which can exist even for equal mass quarks, identifying its location in the phase diagram, the order of the associated phase transitions, and its symmetry breaking pattern, which proves to be the intersection of the symmetry groups of the 2SC and CFL phases.  We also investigate the effects of the strange quark mass on this new phase and the QCD phase diagram generally.  Finally, we analyze the effect of a local color neutrality constraint on these phases of asymmetric pairing.  We observe that for massless quarks the neutrality constraint renders the 2SC phase energetically unfavorable, eliminating it at low temperatures, and giving rise to the previously proposed low temperature critical point, with associated continuity between the hadronic and ACFL phases.  For realistic strange quark masses, however, the neutrality constraint shrinks the 2SC region of the phase diagram, but does not eliminate it, at \textbf{$T = 0$}.

\end{abstract}

\maketitle

\section{Introduction}
The phase structure of strongly interacting matter has seen an explosion of activity in recent years as the boundaries of our experimental probes have continued to expand~\cite{Powell,Callan,Kuti,Pisarski,Fukushima2}.  As facilities such as the Large Hadron Collider and the Relativistic Heavy Ion Collider probe matter of ever higher densities and temperatures, we are able to continually test and refine our theoretical models and understand matter under the extreme conditions encountered in the moments after the big bang and in the cores of neutron stars.

While the fermion sign problem largely restricts the techniques of lattice QCD to zero density, one method for describing strongly interacting matter throughout the phase diagram is the use of effective field theories which are built upon the symmetries of QCD.  One model which has proven useful in this context is the Polyakov--Nambu--Jona-Lasinio (PNJL) model, which was developed to describe dynamical chiral symmetry breaking and has been extended to include quark pairing, confinement, and the QCD axial anomaly~\cite{NJL1,NJL2,Hatsuda2,Buballa,Polyakov,Fukushima,Ratti,Sasaki,Rossner,Abuki,Alford,Kobayashi,tHooft,Abuki_AA,Megias2006,Megias2007}.

An aspect of the QCD phase diagram of particular interest is the nature of quark pairing at intermediate chemical potential, $\mu$.  While it is known that for three quark flavors a color-flavor-locked (CFL) phase, in which all quark flavors and colors pair, is energetically favorable for asymptotically large $\mu$, the preferred pairings for $\mu$ not asymptotically large are not determined.  Calculations indicate phases in which only two colors and flavors pair (2SC)~\cite{Basler}, in which one flavor pairs with all others (uSC, dSC)~\cite{Abuki2006}, and a phase which has properties of both free quarks and hadrons (quarkyonic)~\cite{Fukushima2,McLerran,Hidaka}.

In this paper we build on prior studies of the effects of confinement on quark pairing in the three-flavor PNJL model by considering a wider range of pairing schemes than the CFL and 2SC phases previously considered~\cite{Powell,Basler}.  In particular, by permitting distinct $ud$, $us$, and $ds$ pairing amplitudes, we allow for the possibility that the confining mechanism of QCD may not treat quark flavors, even for equal masses, on an equal footing.  Further, by considering a range of strange quark masses we investigate the combined effects of this potential asymmetry and the decoupling of the strange quark sector with increasing strange quark mass.

We also investigate the implications of a local color neutrality constraint on the phase structure of dense quark matter.  While QCD has the capacity to dynamically achieve local color neutrality by means of a gluon field condensate $\braket{A^0_a}$, the PNJL model lacks the necessary gluonic degrees of freedom to achieve such neutrality in a phase of asymmetric quark pairing (e.g., 2SC, uSC).  Thus, one must impose such neutrality ``by hand" in order to avoid the large color-electric forces which would result from color accumulation~\cite{Gerhold,Dietrich,Iida,Steiner,Buballa2}.  Prior studies of the axial anomaly's influence on the phase structure of dense quark matter in the (P)NJL model have either focused on pairing structures which are trivially color neutral~\cite{Powell,Abuki_AA} or have allowed for locally colored phases~\cite{Basler}.  By introducing an effectively color-dependent chemical potential we impose local color neutrality and study its effects on the low temperature portion of the QCD phase diagram, most notably its suppression of phases of asymmetric quark pairing and the subsequent realization of quark-hadron continuity.

The outline of the paper is as follows.  We begin in Sec.~\ref{sec:model} by recalling the three-flavor PNJL model with axial anomaly.  In Sec.~\ref{sec:results1} we construct the phase diagram for massless QCD and identify a new homogeneous asymmetric color-flavor-locking (ACFL) phase characterized by \textit{breached} pairing in which all quarks pair, but with unequal magnitudes.  In Sec.~\ref{sec:results2} we construct the phase diagram for massive QCD with various strange quark masses in order to study the effects of the strange quark mass on both the ACFL phase and the phase diagram generally.  In Sec.~\ref{sec:NEW} we consider the ACFL phase more carefully by studying the associated phase transitions and symmetry breaking patterns.  Finally, in Sec.~\ref{sec:Neutral} we impose a local color neutrality constraint and investigate the resulting suppression of the 2SC phase at low temperatures.

\section{Three-Flavor PNJL Model \label{sec:model}}
\subsection{Lagrangian}
The Lagrangian for the three-flavor Nambu--Jona-Lasinio model with a Polyakov loop at temperature $T$ is~\cite{Powell,Basler}
\beq
\Lc = \overline{q} (i \slashed{D} - \hat{m} + \mu \gamma^0) q + \Lc^{(4)} + \Lc^{(6)} - \Uc (\Phi,\overline{\Phi},T)   ,   \label{eq:Lagrangian}
\eeq
where the covariant derivative $D_\mu = \partial_\mu - i \delta^0_\mu A_0$ couples a static homogeneous gauge field $A_0$ to the quark field $q$ and $\hat{m}$ is the bare quark mass matrix in flavor space.  $\Lc^{(4)}$ and $\Lc^{(6)}$ are effective four- and six-quark interactions, respectively, and $\Uc(\Phi,\overline{\Phi},T)$ is the Polyakov loop potential, which governs the deconfinement transition in the pure-gauge sector.

The four-quark interaction is invariant under the $\mbox{SU(3)}_L \otimes \mbox{SU(3)}_R \otimes \mbox{U(1)}_B \otimes \mbox{U(1)}_A$ symmetry group of classical QCD, while allowing for spontaneous breaking of chiral symmetry and diquark pairing:
\beq
\Lc^{(4)} = 8 G \mbox{Tr} (\phi^\dagger \phi) + 2 H \mbox{Tr} (d^\dagger_R d_R + d^\dagger_L d_L)  ,   \label{eq:4quark}
\eeq
where $\phi_{ij} = (q_R)^j_a (q_L)^i_a$ is the chiral operator and $(d_R)^i_a = \epsilon_{abc} \epsilon_{ijk} (q_R)^j_b C (q_R)^k_c$ and $(d_L)^i_a = \epsilon_{abc} \epsilon_{ijk} (q_L)^j_b C (q_L)^k_c$ are diquark operators of right and left chirality, respectively, with $C$ the charge-conjugation operator.  The labels $a,b,c$ and $i,j,k$ index color and flavor, respectively.  We take $G,H > 0$, which corresponds to attractive four-quark interactions.

The six-quark interaction reflects the QCD axial anomaly by explicitly breaking $\mbox{U(1)}_A$ while retaining invariance under the remaining (physical) QCD symmetry group:
\beq
\Lc^{(6)} & = & - 8 K \mbox{det} \hspace{.5mm} \phi + K^\prime \mbox{Tr} [ (d^\dagger_R d_L) \phi ] + \mbox{H.c.}
\eeqa

The final ingredient in our model is the Polyakov loop, which serves as an order parameter for confinement in the pure-gauge sector~\cite{Polyakov,Fukushima},
\beq
\Phi (\mathbf{x}) = \frac{1}{3} \hspace{.5mm} \mbox{Tr} \hspace{.5mm} {\cal{P}} \hspace{.5mm} \mbox{exp} \left \{ i \int^\beta_0 d \tau A_0 (\tau, \mathbf{x}) \right \}   ,   \label{eq:Phidef}
\eeq
where $\cal{P}$ is the path-ordering operator and $\beta = 1/T$.  Writing $A_\mu = A^a_\mu \lambda_a / 2$, where the $\lambda_a$ are the Gell-Mann matrices, and working in the Polyakov gauge, in which $A_0$ is diagonal, yields $ A_0 = \phi_3 \lambda_3 + \phi_8 \lambda_8$.

As discussed at length in~\cite{Powell,Rossner}, in order to ensure a real thermodynamic potential we restrict our attention to the case $\phi_8 = 0$.  Making the standard finite-temperature replacements $t \to -i \beta$ and $A_0 \to i A_0$, evaluating Eq. (\ref{eq:Phidef}) explicitly for a homogeneous gauge field yields the relation
\beq
\Phi = \frac{1 + 2 \cos(\beta \phi_3)}{3}   .   \label{eq:Phiphi}
\eeq
Following Fukushima we describe the pure-gauge deconfinement transition via the potential~\cite{Fukushima,Fukushima2}
\beqa
\frac{\Uc}{T^4} & = & - \frac{1}{2} \hspace{.5mm} a(T) \overline{\Phi} \Phi + b(T) \ln [ 1 - 6 \overline{\Phi} \Phi   \nonumber \\
	&& \hspace{25mm} + 4 (\Phi^3 + \overline{\Phi}^3) - 3 (\overline{\Phi} \Phi)^2 ]    ,
\eeqa
where the temperature-dependent coefficients are
\beqa
a(T) = a_0 + a_1 \left(\frac{T_0}{T} \right) + a_2 \left(\frac{T_0}{T} \right)^2   \hspace{2.5mm} , \hspace{2.5mm} b(T) = b_3 \left(\frac{T_0}{T} \right)^3   ,   \nonumber
\eeqa
and the $a_i$ and $b_3$ are chosen to correctly reproduce lattice QCD results (see Table~\ref{tab:polycoefs}).  In addition, while $T_0 = 270$ MeV is the critical temperature for the deconfinement transition in the pure-gauge sector~\cite{Ratti,Sasaki}, when quarks are included in the PNJL model, the transition temperature deviates from $T_0$.  Therefore, in what follows we consider $T_0$ as a parameter of our model, which we will set by matching the deconfinement transition at $\mu = 0$, defined as a maximum in $d \Phi / dT$ (as discussed in~\cite{Aoki1,Aoki2}), to the lattice QCD value of $T^{QGP}_c = 176$ MeV.

\begin{center}
\begin{table}
\caption{\footnotesize{Coefficients of the Polyakov-loop potential~\cite{Rossner}.}}
\begin{tabular}{p{20mm}p{20mm}p{20mm}p{20mm}}
\hline \hline
$\hspace{1mm}a_0$ & \hspace{3mm}$a_1$ & \hspace{7mm}$a_2$ & \hspace{15mm}$b_3$    \\
\hline
3.51 &  -2.47 &  \hspace{5.5mm}15.2 &  \hspace{12mm}-1.75  \\
\hline \hline
\end{tabular}
\label{tab:polycoefs}
\end{table}
\end{center}

\vspace{-10mm}
\subsection{Thermodynamic potential}
Working in the mean field, we consider the homogeneous scalar chiral and diquark condensates
\beq
\braket{\overline{q}^i_a q^j_a} = \sigma_i \delta_{ij} \hspace{5mm} , \hspace{5mm} \braket{q^T C \gamma_5 t_i l_j q} = d_i \delta_{ij}   .   \label{eq:conds}
\eeq
Note that there is no sum over $i$; rather, the right sides of Eq. (\ref{eq:conds}) are diagonal matrices in flavor space with three distinct elements.  As shown in~\cite{Powell,Basler} the mean field Lagrangian becomes
\beqa
\Lc_{MF} & = & \sum^3_{j=1} \overline{q}_j \left(i \slashed{\partial} - M_j + (\mu + i \phi_3 \lambda_3) \gamma^0 \right) q_j   \nonumber \\
	 && \hspace{5mm} - \frac{1}{2} \sum^3_{j=1} [ \Delta^\ast_j (q^T C \gamma_5 t_j l_j q) + \mbox{H.c.} ]   \label{eq:MFLag}\\
	 && \hspace{5mm} - {\cal{V}} - \Uc   ,   \nonumber
\eeqa
where the $t_j$ and $l_j$ are the antisymmetric Gell-Mann matrices in flavor and color space respectively, ${\cal{V}}$ is given explicitly below, the effective mass of the $j$th quark flavor is
\beq
M_j = m_j - 4 G \sigma_j + K |\epsilon_{jkl}| \sigma_k \sigma_l + \frac{K^\prime}{4} \hspace{.5mm} |d_j|^2   ,
\eeq
and the $j$th pairing gap is
\beq
\Delta_j = - 2 \left(H - \frac{K^\prime}{4} \hspace{.5mm} \sigma_j \right) d_j   .
\eeq
We choose $(u,d,s)$ and $(r,g,b)$ as flavor and color bases and use the Gell-Mann matrices $\lambda_{7,5,2}$ as a representation of $t_{1,2,3}$ and $l_{1,2,3}$, so that $\Delta_1$ represents the $d_g s_b$ and $d_b s_g$ pairing gap, while $\Delta_2$ and $\Delta_3$ represent the gaps of $u_r s_b$ and $u_b s_r$ and $u_r d_g$ and $u_g d_r$ pairs respectively.

Introducing the Nambu-Gor'kov spinor $\Psi = (q \hspace{1mm} C\overline{q}^T)^T / \sqrt{2}$, we may recast our model as a free theory with $\Lc = \overline{\Psi} S^{-1} \Psi - V - \Uc$, where the inverse propagator in Nambu-Gor'kov and momentum space is
\beq
S^{-1} (k) = \left(\begin{matrix}
			\slashed{k} - \hat{M} + \mu^\prime \gamma^0 & \Delta_j \gamma_5 t_j l_j \\
			- \Delta^\ast_j \gamma_5 t_j l_j & \slashed{k} - \hat{M} - \mu^\prime \gamma^0
		   \end{matrix} \right)   ,   \label{eq:invprop}
\eeq
with $\mu^\prime = \mu + i \phi_3 \lambda_3$ and where the sum over $j$ in the off-diagonal elements is implied.  The condensates directly contribute a potential
\beqa
{\cal{V}} & = & 2 G \sum^3_{j=1} \sigma^2_j - 4 K \sigma_1 \sigma_2 \sigma_3 + \sum^3_{j=1} \left(H - \frac{K^\prime}{2} \hspace{.5mm} \sigma_j \right) |d_j|^2   .   \nonumber \\
\eeqa

Integrating over the Nambu-Gor'kov fields and performing the resulting Matsubara sum yields the thermodynamic potential
\beq
\Omega & = & - \frac{T}{2} \sum^{72}_{n=1} \int^\Lambda \frac{d^3 \mathbf{k}}{(2\pi)^3} \left[ \ln (1 + e^{-\beta E_n}) + \frac{1}{2} \hspace{.5mm} \beta \Delta E_n \right]   \nonumber \\   
         && \hspace{10mm} + {\cal{V}} + \Uc   ,   \label{eq:Omega}
\eeq
where the $E_n$ are the 72 poles of the inverse propagator in Eq.~(\ref{eq:invprop}), $\Delta E_n = E_n - E^{free}_n$ is the difference between the eigenvalue and its noninteracting value (without the absolute value), and the factor of $1/2$ accounts for the double counting of degrees of freedom in the Nambu-Gor'kov formalism.

Note that in Eq.~(\ref{eq:Omega}) we have introduced a high-momentum cutoff $\Lambda$ to regulate the integral.  The value of $\Lambda$, along with the coupling constants $G$ and $K$, is initially fit to empirical mesonic properties and is given in Table~\ref{tab:couplings} (parameter set I).  Following Abuki \textit{et al.}, as we adjust the strange quark mass and coupling $K^\prime$ (parameter sets II-IX), rather than recalculating $G$ by again fitting the mesonic quantities, we instead, for the sake of simplicity, choose $G$ to yield a fixed value for $(M_u + M_d)/2 = 367.5$ MeV~\cite{Abuki_AA}.  The quantitative effects of this choice are negligible for the present purposes.

Finally, we consider two values of $K^\prime$: (1) $K^\prime = K$, which is suggested by applying a Fierz transformation to the instanton vertex~\cite{Abuki_AA}, and (2) $K^\prime = 4.2 K$, which allows the realization of a low $T$ critical point and provides for easy comparison to the current literature~\cite{Abuki_AA,Basler}.  The remaining couplings are $H \Lambda^2 = 1.74$ and $K \Lambda^5 = 12.36$~\cite{Ratti,Rossner,Abuki_AA}.

\begin{center}
\begin{table}[t]
\caption{\footnotesize{Parameter sets for the three-flavor PNJL model: the strange quark bare mass $m_s$, coupling constants $G$ and $K^\prime$, and Polyakov loop parameter $T_0$, with a spatial momentum cutoff $\Lambda=602.3$ MeV \cite{Buballa}.  Also shown is the constituent strange quark mass at $\mu = T = 0$.  $^\ast$In parameter set I all bare quark masses are set to zero.  In all others we take $m_u = 2.5$ MeV and $m_d = 5.0$ MeV~\cite{PDG}.}}
\begin{tabular}{|c|c|c|c|c|c|c|}
\hline 
& $m_s$ (MeV) & \ $G \Lambda^2$ \ & \ $K^\prime \Lambda^5$ \ & $T_0$ (MeV) \ & $M_s$ (MeV) \\ \hline \hline
\ I$^\ast$ \ & 0 & 1.926 & 12.36 & 210 & 355.1 \\ \hline
\ II \ & 5 & 1.928 & 12.36 & 208 & 369.4 \\ \hline
\ III \ & 5 & 1.928 & 51.91 & 208 & 369.4 \\  \hline
\ IV \ & 20 & 1.915 & 12.36 & 207 & 392.2 \\  \hline
\ V \ & 20 & 1.915 & 51.91 & 207 & 392.2 \\  \hline
\ VI \ & 40 & 1.899 & 12.36 & 206 & 417.5 \\  \hline
\ VII \ & 40 & 1.899 & 51.91 & 206 & 417.5 \\  \hline
\ VIII \ & 80 & 1.877 & 12.36 & 204 & 476.6 \\  \hline
\ IX \ & 80 & 1.877 & 51.91 & 204 & 476.6 \\  \hline
\end{tabular}
\label{tab:couplings}
\end{table}
\end{center}

\vspace{-10mm}
In the next two sections we construct the phase diagram of three-flavor QCD, first in the limit of massless quarks, and then with realistic quark masses.  In order to facilitate a comparison with the current literature, which largely ignores the complication of a local color neutrality constraint, we begin by constructing the phase diagrams without enforcing color neutrality, deferring a discussion of the effects of color neutrality to Sec.~\ref{sec:Neutral}.  We also note that while a variety of spatially inhomogeneous phases (e.g., crystalline color superconductors, Fulde-Ferrell-Larkin-Ovchinnikov phases) may be energetically preferred in certain high density regions of the phase diagram~\cite{Bedaque2003,Caldas2004}, in this paper we consider only homogeneous phases.

\section{Massless QCD Phase Diagram \label{sec:results1}}
\subsection{Without confinement}
In this section we discuss the phase structure of massless QCD before moving on to consider the case of three different mass quarks.  This will allow us to investigate both the general effects of quark mass on the phase diagram and its particular influence on a possible ACFL phase.  Note we do not impose color neutrality in either this section or the following, so that we may consider the effects of this additional constraint in Sec.~\ref{sec:Neutral}.

\begin{figure}
\includegraphics[scale=1.0]{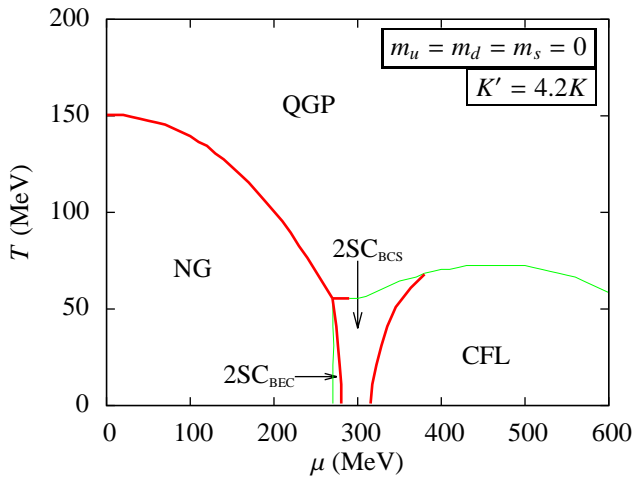}
\caption{\footnotesize{(color online).  Phase diagram for the NJL model (no confinement) with three massless quark flavors.  Thick (red) lines denote first-order transitions while thin (green) lines denote second-order transitions.  The first-order 2SC$_{\mbox{\tiny{BEC}}}$-2SC$_{\mbox{\tiny{BCS}}}$ transition is defined by $M_{u,d}(\mu,T) = \mu$.}}
\label{fig:NJLpd}
\end{figure}

When we turn off confinement by setting $\phi_3 = 0$ and dropping the potential $\Uc (\phi_3, T)$, the thermodynamic potential reduces to that considered by Basler and Buballa~\cite{Basler}.  The only significant difference between the massless NJL model considered here and the massive case is that for massless quarks the chiral phase transition is first-order for all $\mu$, rather than a smooth crossover at low $\mu$ (Fig.~\ref{fig:NJLpd}).  Basler and Buballa have shown that for $K^\prime \gtrsim 3.5 K$ a 2SC$_{\mbox{\tiny{BEC}}}$ phase appears, which we define as a 2SC phase ($d_1 = d_2 = 0$, $d_3 \neq 0$) of diquark pairs in the strongly bound Bose-Einstein condensate (BEC) regime, in which $M_{u,d} > \mu$.  This phase is similarly visible in Fig~\ref{fig:NJLpd}, separated from the chirally broken Nambu-Goldstone (NG) phase by a second-order phase transition, and from a 2SC$_{\mbox{\tiny{BCS}}}$ phase of weakly bound Bardeen-Cooper-Schrieffer (BCS) quark pairs (in which $M_{u,d} < \mu$) by a first-order transition.

\begin{figure}
\includegraphics[scale=1.0]{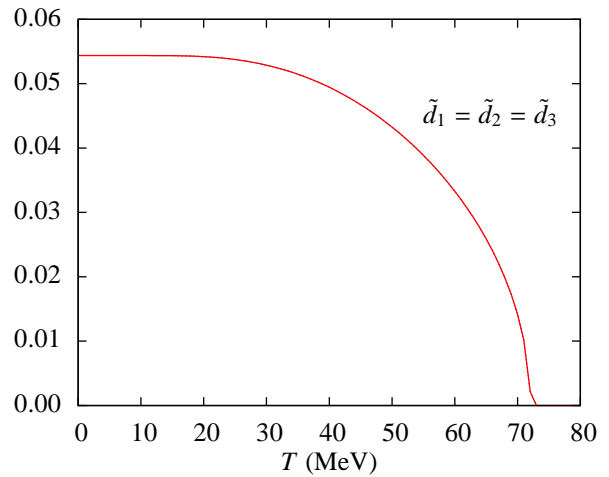}
\caption{\footnotesize{(color online).  Dimensionless diquark condensate $\tilde{d}_i = d_i / \Lambda^3$ in the non-confining massless NJL model for $\mu = 500$ MeV.  As indicated in Fig.~\ref{fig:NJLpd}, the system undergoes a second-order phase transition from the CFL phase to the QGP at 73 MeV.  The linear approach to zero for $ 71 \mbox{ MeV} < T < 73 \mbox{ MeV}$ is due to an effective $\sigma |d|^2$ coupling~\cite{Powell,Hatsuda1,Yamamoto}.}}
\label{fig:NJLdvsT}
\end{figure}

Anticipating the ACFL phase discussed in Secs.~\ref{sec:results2} and~\ref{sec:NEW}, in Fig.~\ref{fig:NJLdvsT} we show the single diquark condensate of the CFL phase as a function of temperature for $\mu = 500$ MeV.  We note that it is roughly constant for $T \lesssim 30$ MeV, and then falls as $d \sim \sqrt{T_c - T}$ for $30 \mbox{ MeV} \lesssim T < 71$ MeV, before finally vanishing as $d \sim (T_c - T)$, due to the effective $\sigma |d|^2$ coupling induced by the axial anomaly~\cite{Powell,Hatsuda1,Yamamoto}.

\subsection{With confinement}
\begin{figure}
\includegraphics[scale=1.0]{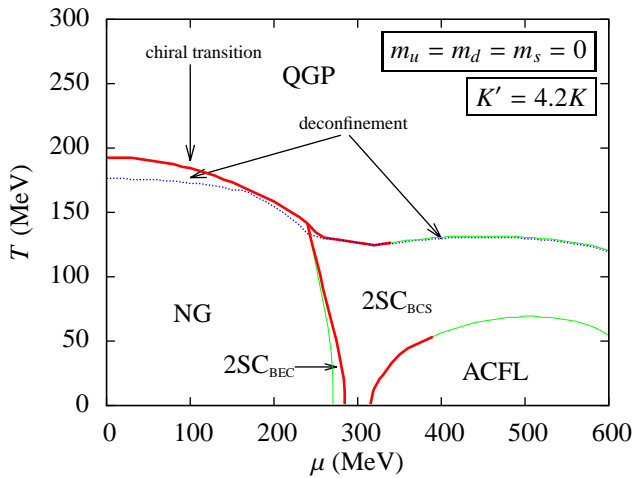}
\caption{\footnotesize{(color online).  Phase diagram for the PNJL model with three massless quark flavors.  Line types have the same meaning as in Fig.~\ref{fig:NJLpd}, with the additional dotted (blue) line representing the deconfinement crossover.}}
\label{fig:PNJLpd}
\end{figure}

In order to construct the phase diagram in the presence of the Polyakov loop we first fix $T_0$ by matching the model's deconfinement transition at $\mu = 0$ to the lattice value of $T^{QGP}_c = 176$ MeV.  The resulting value of $T_0$ varies slightly with $m_s$, and is given for the various parameter sets used in Table~\ref{tab:couplings}.  Minimizing $\Omega$ with respect to the condensates and Polyakov loop, we obtain the phase diagram shown in Fig.~\ref{fig:PNJLpd}.  As has been widely reported, the inclusion of the Polyakov loop pulls the chiral transition to higher temperatures (from 151 MeV to 193 MeV), significantly enlarging the region of symmetry breaking~\cite{Ratti,Rossner,Enlarged}.

One important consequence of the increase of $T^{\mbox{\tiny{QGP}}}_c$ is that the Polyakov loop gives rise to a much larger region of 2SC$_{\mbox{\tiny{BCS}}}$, which we define as a 2SC phase ($d_1 = d_2 = 0$, $d_3 \neq 0$) in which $M_{u,d} < \mu$.  In particular, this phase now persists to much higher $\mu$ than in the NJL model, where it is constrained to roughly $270 \mbox{ MeV} \lesssim \mu \lesssim 350$ MeV.

\begin{figure}
\includegraphics[scale=1.0]{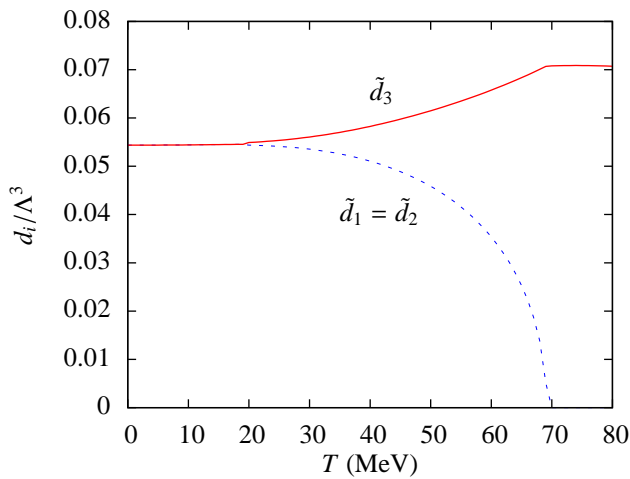}
\caption{\footnotesize{(color online).  Dimensionless diquark condensates in the massless PNJL model for $\mu = 500$ MeV.}}
\label{fig:PNJLdvsT}
\end{figure}

Figure~\ref{fig:PNJLdvsT} shows the two distinct diquark condensates $d_1 = d_2$ and $d_3$ for $\mu = 500$ MeV.  We find that for $T \lesssim 20$ MeV, the results are not significantly altered from the NJL model.  However, for $T \gtrsim 20$ MeV we find that $d_1 = d_2$ falls with increasing $T$, while $d_3$ increases until the system undergoes a second-order phase transition to the 2SC$_{\mbox{\tiny{BCS}}}$ phase at 70 MeV, slightly below the location of the transition to the quark-gluon plasma (QGP) in the absence of confinement.  Thus, the ground state of the system at intermediate $\mu$ is no longer a symmetric CFL phase, but rather an asymmetric CFL phase characterized by $0 < d_1 = d_2 < d_3$.

\begin{figure*}
        \begin{subfigure}[b]{0.45\textwidth}
                \centering
                \includegraphics[width=\textwidth]{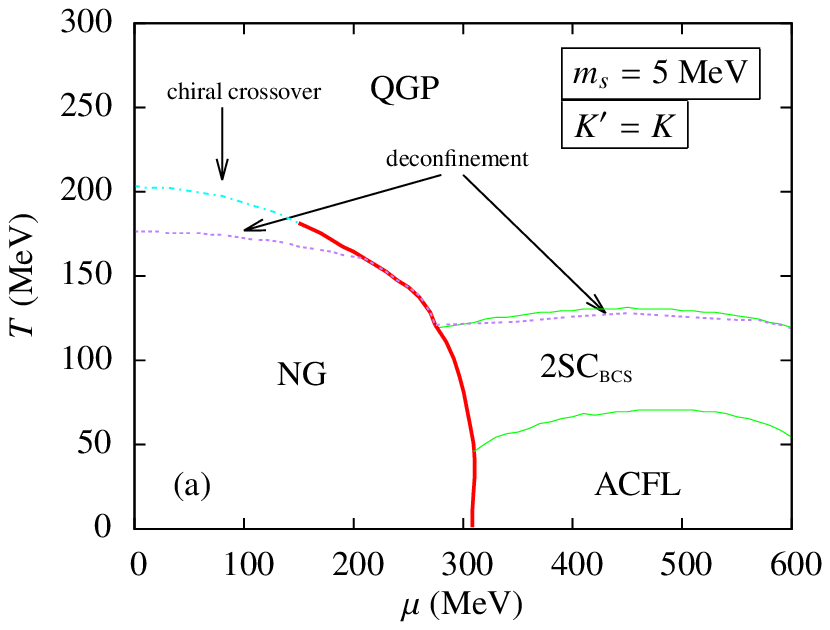}
        \end{subfigure}
        \begin{subfigure}[b]{0.45\textwidth}
                \centering
                \includegraphics[width=\textwidth]{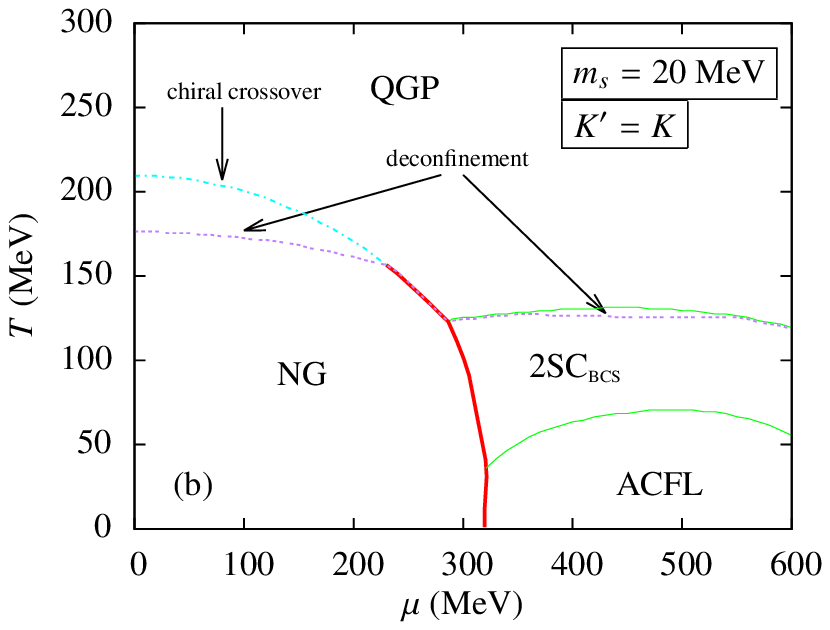}
        \end{subfigure}
        \begin{subfigure}[b]{0.45\textwidth}
                \centering
                \includegraphics[width=\textwidth]{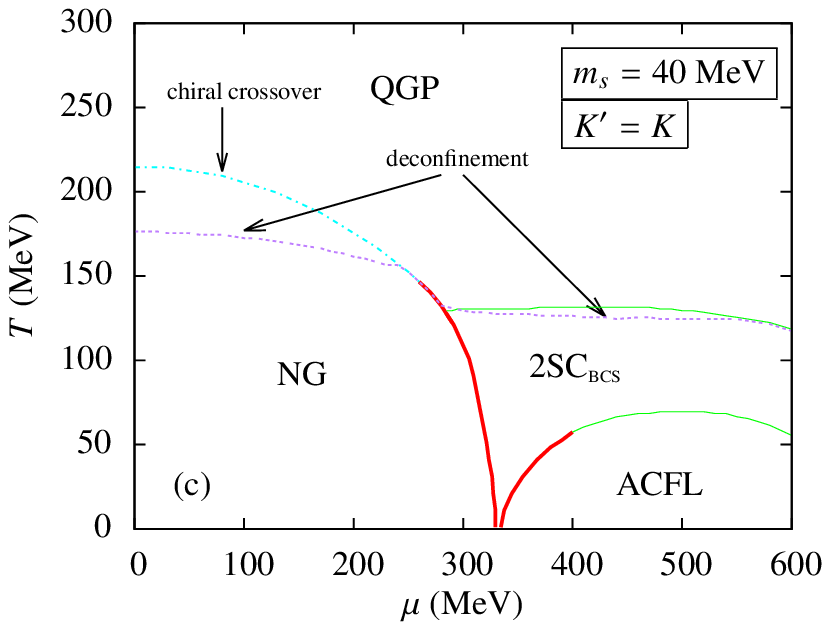}
        \end{subfigure}
        \begin{subfigure}[b]{0.45\textwidth}
                \centering
                \includegraphics[width=\textwidth]{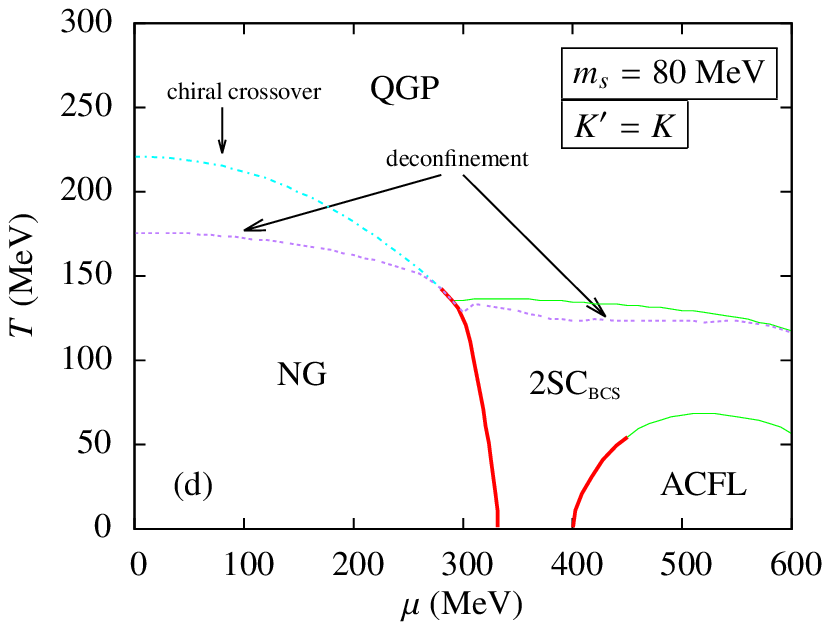}
        \end{subfigure}
        \caption{(color online).  Phase diagrams for the PNJL model with realistic up and down quark masses and various strange quark masses, where the axial anomaly couplings are taken to be equal ($K^\prime = K$).  Line types have the same meaning as in Fig.~\ref{fig:PNJLpd}.}
	\label{fig:Kp1}
\end{figure*}

\section{Realistic Mass QCD Phase Diagram \label{sec:results2}}

Having observed the emergence of an ACFL phase in massless QCD, we now consider the effects of realistic bare quark masses on both this phase and the phase diagram generally.  To do so we construct phase diagrams for $m_s = 0$, 20, 40, and 80 MeV.  In all cases we take $m_u = 2.5$ MeV and $m_d = 5.0$ MeV, while the coupling $G$ is adjusted in order to maintain $(M_u + M_d)/2 = 367.5$ MeV at $\mu = T = 0$.

\begin{figure*}
        \begin{subfigure}[b]{0.45\textwidth}
                \centering
                \includegraphics[width=\textwidth]{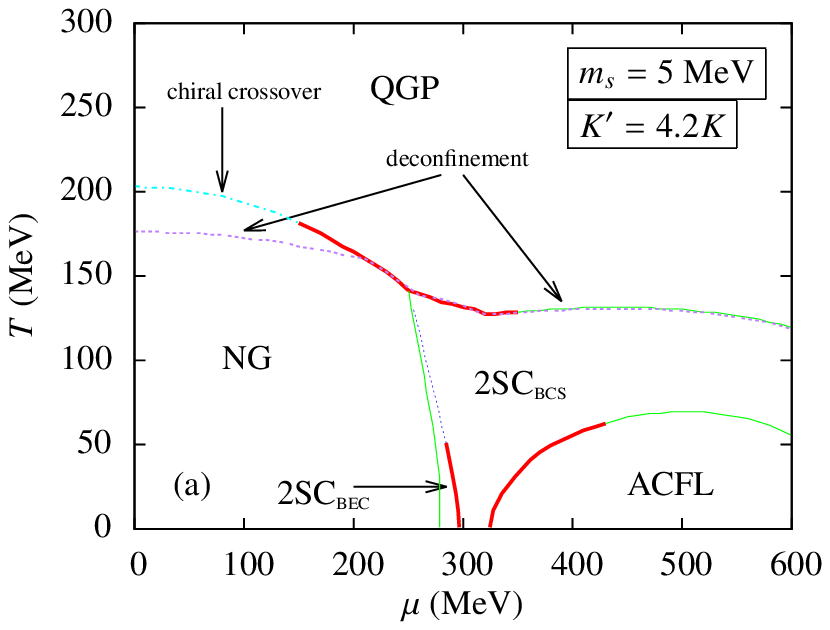}
        \end{subfigure}
        \begin{subfigure}[b]{0.45\textwidth}
                \centering
                \includegraphics[width=\textwidth]{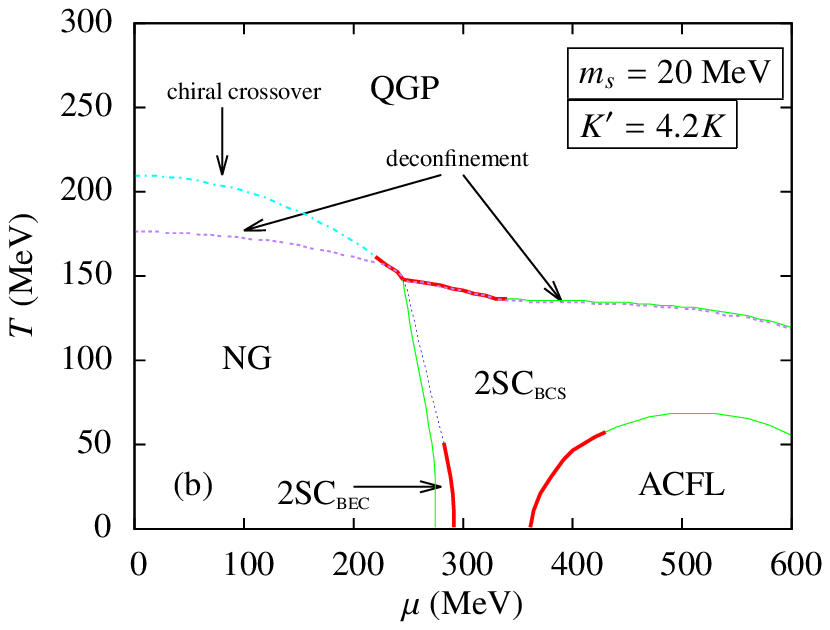}
        \end{subfigure}
        \begin{subfigure}[b]{0.45\textwidth}
                \centering
                \includegraphics[width=\textwidth]{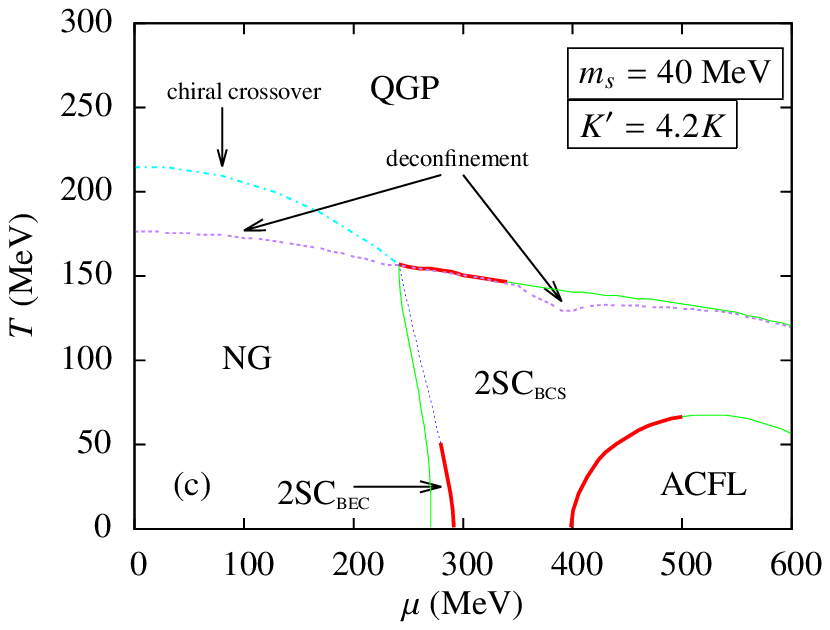}
        \end{subfigure}
        \begin{subfigure}[b]{0.45\textwidth}
                \centering
                \includegraphics[width=\textwidth]{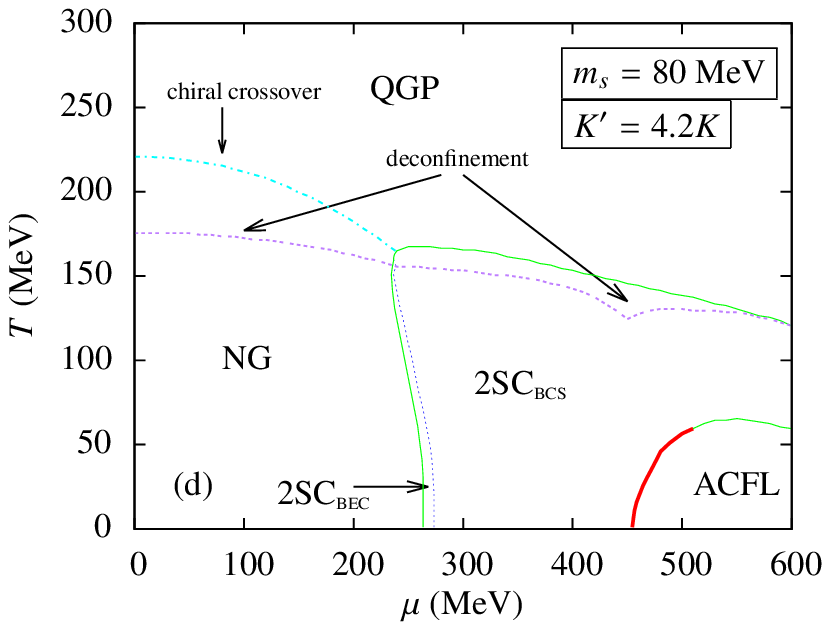}
        \end{subfigure}
        \caption{(color online).  Phase diagrams for the PNJL model with realistic up and down quark masses and various strange quark masses, where $K^\prime = 4.2 K$.  Line types have the same meaning as in Fig.~\ref{fig:PNJLpd}.}
	\label{fig:Kp42}
\end{figure*}

As shown in Figs.~\ref{fig:Kp1} and~\ref{fig:Kp42}, as the strange quark mass increases, the region of ACFL moves to higher $\mu$, effectively decoupling the strange quark from the up and down sector.  This is due to the fact that in the limit $m_s \to \infty$, there is insufficient energy to generate strange quarks and we are left with an effectively two-flavor system.  We also note that while for small $m_s$ the deconfinement transition at large $\mu$ essentially coincides with the breaking of up and down quark Cooper pairs (the 2SC-QGP transition), as $m_s$ increases the deconfinement temperature moves down somewhat.  

Also noteworthy is the fact that except for small $m_s$ and $K^\prime$, a critical point appears on the ACFL-2SC phase boundary, separating a first-order transition at lower $\mu$ from a second-order transition at higher $\mu$.  One can summarize the situation by noting that the ACFL-2SC transition is first-order when $T_c \lesssim 50$ MeV, and second-order when $T_c \gtrsim 50$ MeV.  Thus, for example, for $m_s = 5, 20$ MeV and $K^\prime = K$, the phase boundary never drops below $T \approx 50$ MeV and the transition is always second-order.  We note, however, that while the phase boundary has a negative slope for large $\mu$, the transition does not again become first-order when the boundary drops below $T \approx 50$ MeV.

As shown by Abuki \textit{et al.} for the nonconfining NJL model and by the present authors for the massless PNJL model, we find that for $K^\prime \geq 4.2 K$, a low $T$ critical point emerges~\cite{Powell,Abuki_AA}.  Also, as shown by Basler and Buballa, when one allows for 2SC pairing this critical point acts as the termination of a line of first-order BEC-BCS transitions, above which a smooth crossover develops~\cite{Basler}.  Interestingly, as shown in Fig.~\ref{fig:Kp42}, when the 2SC$_{\mbox{\tiny{BEC}}}$ phase exists, we find that for $m_s = 5$, 20, and 40 MeV the BEC-BCS transition is first-order at zero temperature, while for $m_s = 80$ MeV the critical point drops below the $T$ axis and one obtains a smooth BEC-BCS crossover.

\begin{figure}
\includegraphics[scale=1.0]{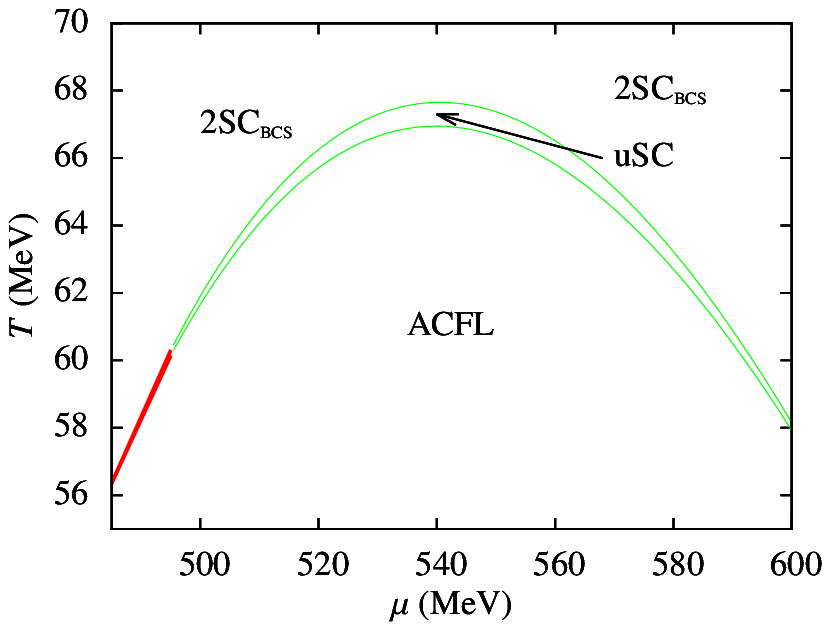}
\caption{\footnotesize{(color online).  Partial phase diagram of the three flavor PNJL model with $K^\prime = 4.2 K$ and (bare) masses of $m_u = 0$, $m_d = 40$ MeV, and $m_s = 80$ MeV.  Line types have the same meaning as in Fig.~\ref{fig:NJLpd}.}}
\label{fig:zoom}
\end{figure}

While not visible in Figs.~\ref{fig:Kp1} and~\ref{fig:Kp42}, for unequal mass quarks much of the ACFL-2SC phase boundary is actually two distinct, but very closely spaced phase boundaries.  The first boundary, at slightly lower temperatures, separates the ACFL phase from a sliver of a uSC phase in which up/down and up/strange quarks pair, but down/strange quarks do not.  Thus, crossing this phase boundary corresponds to breaking the down/strange quark Cooper pairs.  The second boundary separates the uSC phase from the 2SC and corresponds to the breaking of the up/strange quark pairs.  Figure~\ref{fig:zoom} shows these two distinct transitions for exaggerated up and down quark masses ($m_u = 0$, $m_d = 40$ MeV, $m_s = 80$ MeV), in order to make the distinct phase boundaries visible.

We note that while the precise value of $K^\prime$ is unknown, on the basis of the Fierz transformation mentioned in Sec.~\ref{sec:model} it is expected that $K^\prime \sim K$, and it is unclear if any mechanism might increase $K^\prime$ above the $4.2 K$ threshold required to realize the low temperature critical point and BEC-BCS crossover.  It seems more likely that Fig.~\ref{fig:Kp1}(d) is closest to the true QCD phase diagram.

Finally, a word is required regarding quark pairing in the ACFL phase for realistic quark masses.  While the splitting of up and down quark masses is quite small relative to the chemical potential at which the ACFL phase is obtained ($\mu \sim 400$ MeV), the mass splitting between the strange quark and the two light flavors is indeed large ($M_s - M_{u,d} \gtrsim 100$ MeV).  This mass difference results in significantly mismatched Fermi surfaces, which act as a barrier to quark pairing in the conventional BCS picture of superconductivity.  However, with the assumption of spatially uniform pairing, the different dispersion relations of the ultrarelativistic light quarks on the one hand, and the much slower strange quarks on the other, can lead to a situation in which quarks on the strange quark Fermi surface pair with quarks on the \textit{interior} of the light flavors' Fermi spheres, as shown in~\cite{Gubankova2003}.  This \textit{breached} pairing corresponds to a situation in which the $T = 0$ state of the system consists of both superfluid and normal Fermi liquid components with both gapped and ungapped quasiparticle excitations~\cite{Liu2004}.  Thus, as shown in~\cite{Forbes2005}, it is indeed possible to form a stable homogeneous superfluid phase out of the mismatched Fermi spheres, as we observe.

\section{ACFL Phase \label{sec:NEW}}
\subsection{Quark pairing amplitudes}
\begin{figure}
\includegraphics[scale=1.0]{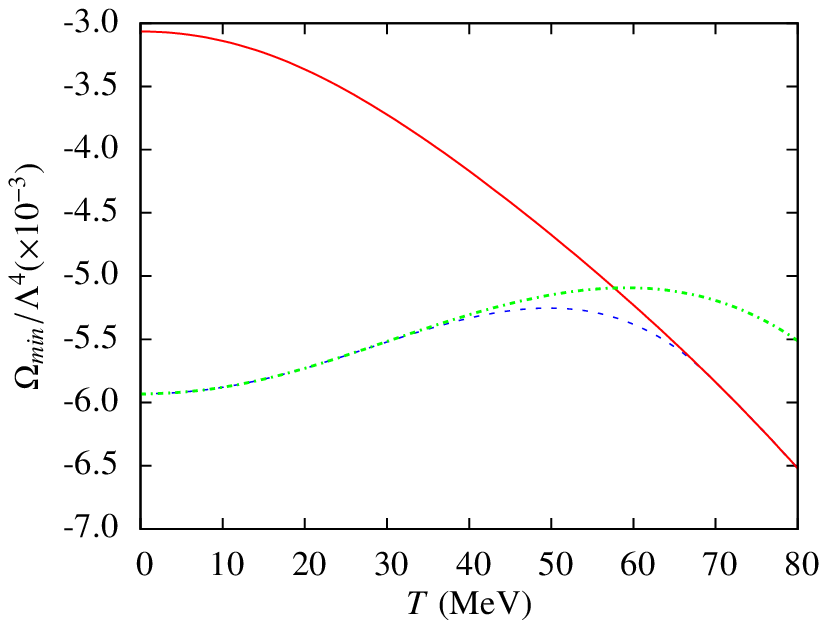}
\caption{\footnotesize{(color online).  Minimum $\Omega / \Lambda^4$ vs. $T$ at $\mu = 500$ MeV for the phases: CFL (dot-dash; green), ACFL (dotted; blue), and QGP (solid; red) in the massless PNJL model.}}
\label{fig:phases}
\end{figure}

The evolution of color superconducting quark matter with increasing temperature can be inferred from Fig.~\ref{fig:phases}.  At low temperatures, the ACFL phase is essentially identical to the CFL phase, with $d_1 = d_2 \approx d_3$, and has a thermodynamic potential well below the QGP.  At high temperatures, the ACFL phase morphs continuously into the 2SC$_{\mbox{\tiny{BCS}}}$ phase, with $d_1 = d_2 = 0$, via a second-order phase transition.  In between these two limiting cases, for $20 \mbox{ MeV} < T < 70 \mbox { MeV}$, the ACFL phase is distinct from both the 2SC and QGP phases, and has a thermodynamic potential below both.

We also note that while not clearly visible in Fig.~\ref{fig:PNJLdvsT}, our calculations indicate that for $T > 6$ MeV it is always energetically favorable to adopt unequal pairing amplitudes.  Thus, while we cannot exclude the possibility of a low temperature CFL-ACFL phase transition, it seems very likely that the unequal pairing amplitudes persist to arbitrarily low temperatures, and that a symmetric CFL phase at intermediate $\mu$ is restricted to $T = 0$.

We can understand the asymmetric behavior of the quark pairing by noting that in our chosen gauge (and with $\phi_8 = 0$) the quark-Polyakov loop coupling is of the form
\beq
\overline{q} A_0 \gamma^0 q = \phi_3 ( \overline{r} \gamma^0 r - \overline{g} \gamma^0 g)   ,
\eeq
where we have written the color indices explicitly, so that the Polyakov loop couples only to red and green quarks.  Thus, the condensates $d_1$ (which involves green and blue quarks) and $d_2$ (red and blue) are only singly coupled to the Polyakov loop, while $d_3$ (green and red) is doubly coupled.

One may inquire whether this phase of unequal quark pairing is simply an artifact of our choice of $\phi_8 = 0$, or whether such a phase might actually be realized in QCD.  Unfortunately, in the present model, allowing $\phi_8 \neq 0$ renders the thermodynamic potential complex so that its minimization is no longer a well-posed problem.  Nevertheless, our results do demonstrate the possibility of obtaining a phase characterized by unequal quark pairing, and they present a challenge to other models of dense quark matter to address the question of its realization.

In addition to local color charge, asymmetric quark pairing in both the ACFL and 2SC phases can give rise to a net local \textit{electric} charge. In quark matter in neutron stars such a charge is neutralized by a net electron (and possibly muon) density, and indeed in deriving an equation of state for neutron stars, we must include charge neutrality.  On the other hand, matter encountered in heavy ion collisions is electrically charged, and the collisions occur on sufficiently short time scales so that while the matter reaches equilibrium with respect to the strong nuclear force, it does not reach charge equilibrium.  While we discuss color neutrality in Sec.~\ref{sec:Neutral}, we do not further consider electrical neutrality in this paper.

\subsection{Symmetry breaking pattern}
Having identified the region of the phase diagram occupied by the ACFL phase as well as the order of the associated phase transitions, we next study the symmetry breaking pattern of this phase.  We begin by noting that the symmetry groups of the 2SC and CFL states are~\cite{Alford,Alford2}
\beq
& \mbox{2SC}: & \mbox{SU(2)}_{rg} \otimes \mbox{SU(2)}_L \otimes \mbox{SU(2)}_R \otimes \mbox{U(1)}_{\bar{B}} \otimes \mbox{U(1)}_S   ,   \nonumber \\
& \mbox{CFL}: & \mbox{SU(3)}_{c+L+R} \otimes Z_2   \nonumber   ,
\eeq
where SU(2)$_{rg}$ denotes a rotation in the color subspace of red and green quarks, U(1)$_{\bar{B}}$ is a rotated baryon conserving symmetry with conserved quantity
\beq
\overline{B} = \overline{Q} + I_3   \hspace{5mm}   \mbox{with}   \hspace{5mm}   \overline{Q} = Q - \frac{1}{2 \sqrt{3}} \hspace{.5mm} \lambda_8   ,
\eeq
where $I_3$ is the isospin operator, $Q$ and $\overline{Q}$ are the standard and rotated (conserved) electromagnetic charge operators in the 2SC phase, and U(1)$_S$ corresponds to multiplying the strange quark by an arbitrary phase.

Since both the CFL and 2SC phases are special cases of the ACFL phase, the symmetry group of the ACFL phase must be a subset of the symmetry groups of these respective phases.  Thus, the color-flavor-locking aspect of the CFL phase requires that there be no unbroken independent color or chiral rotations in the ACFL phase, while the SU(2)$_{rg}$ symmetry of the 2SC phase requires that there be no unbroken symmetry which mixes blue quarks with red or green quarks.  A direct calculation demonstrates that none of the remaining symmetries is broken and we are left with the symmetry group
\beq
\mbox{ACFL}: \mbox{SU(2)}_{rg+L+R} \otimes Z_2   .   \nonumber
\eeq

In fact, the symmetry group of the ACFL phase is simply the intersection of the symmetry groups of the 2SC and CFL phases.  Moreover, this symmetry group is identical to that of the CFL phase with unequal strange quark mass~\cite{symmetryfoot}.  Finally, we note that we expect 14 Goldstone bosons in the ACFL phase, which follows from the $8_R + 8_L + 1_B = 17$ generators of the Lagrangian and the three generators of the ACFL symmetry group.

\section{Color Neutrality \label{sec:Neutral}}

In the prior sections we have constructed the phase diagram of the PNJL model for both massless and massive quarks and have observed the emergence of a new ACFL phase at large $\mu$.  If our model is to accurately reflect the behavior of dense QCD, however, for the homogeneous phases which we consider here we must also investigate the effects of the requirement of local color neutrality.  In fact, both the 2SC phase previously reported by Basler and Buballa~\cite{Basler} and the new ACFL phase possess nonzero color densities which would, if left unchecked, induce large color-electric forces in the superconducting quark matter.

The origin of the net color density, in both the 2SC and ACFL phases, is the modification of the quark dispersion relations which results from unequal pairing amplitudes for red and green quarks compared with blue quarks.  In the 2SC phase, for example, at fixed particle number the pairing of red and green quarks results in a decrease in the Fermi energy of these colors.  In a system at fixed quark chemical potential $\mu$, this results in an increase in the density of red and green quarks compared to the unpaired blue quarks, and a corresponding net antiblue color density.  In QCD, this quark color density is exactly canceled by the development of a nonzero expectation value of the gluon field (i.e., tadpole diagrams), and so the homogeneous 2SC phase remains color neutral~\cite{Gerhold,Dietrich}.  However, having replaced the local SU(3) color symmetry of QCD with the global symmetry of the PNJL model we now lack the means for dynamically realizing a neutral ground state.

\begin{figure*}
        \begin{subfigure}[b]{0.45\textwidth}
                \centering
                \includegraphics[width=\textwidth]{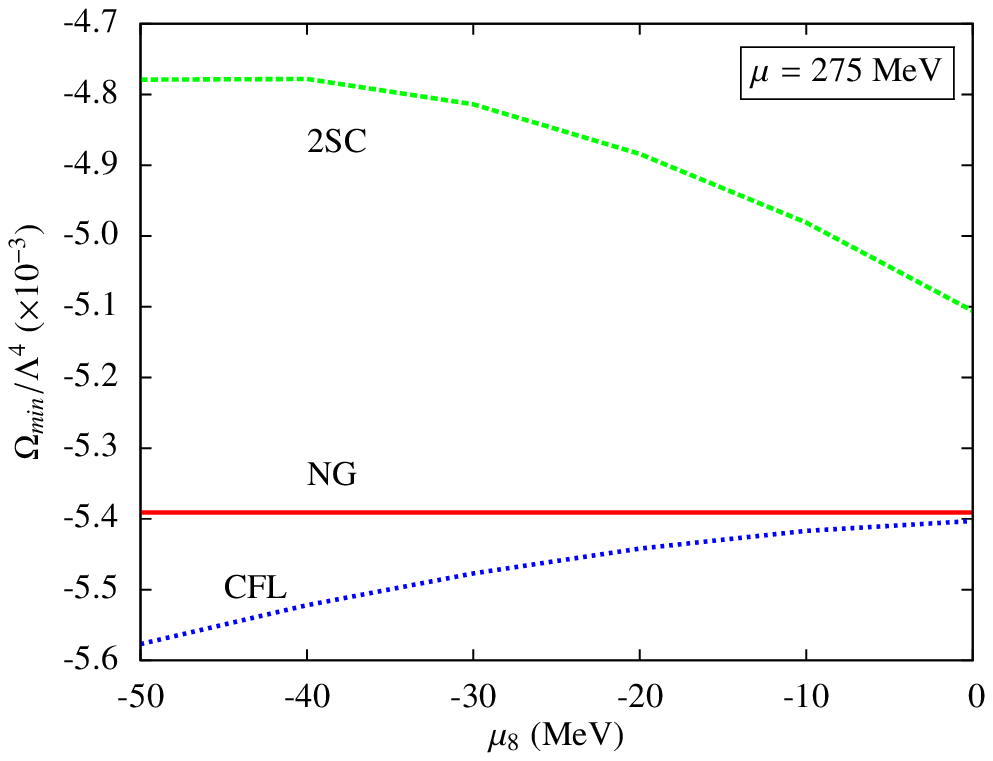}
	\end{subfigure}
        \begin{subfigure}[b]{0.45\textwidth}
                \centering
                \includegraphics[width=\textwidth]{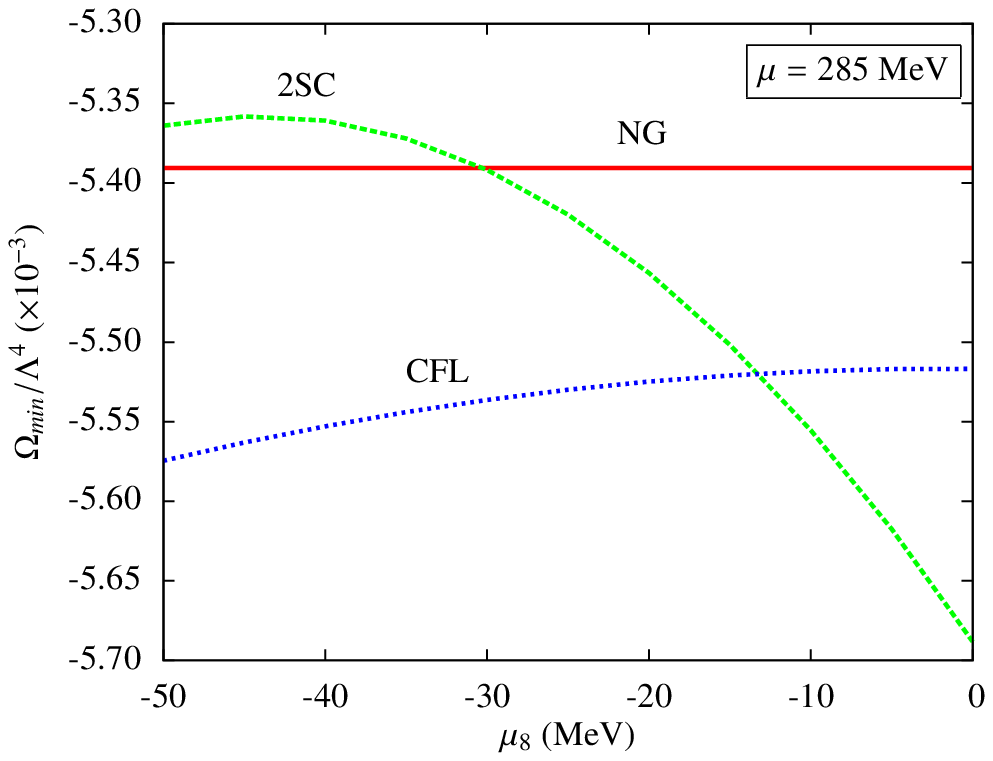}
	\end{subfigure}
\caption{(color online).  Minimum thermodynamic potentials in the NG, 2SC, and CFL phases as a function of $\mu_8$ at $T = 0$ for $\mu = 275$ and $285$ MeV, in the three-flavor PNJL model with massless quarks and $K^\prime = 4.2 K$.}
\label{fig:omegas}
\end{figure*}

The standard method for imposing color neutrality in the NJL model is to introduce a set of color chemical potentials $\mu_a$, which are chosen to ensure vanishing color densities~\cite{Iida,Steiner,Buballa2}:
\beq
n_a = \braket{q^\dagger T_a q} = - \frac{\partial \Omega}{\partial \mu_a} = 0   ,
\eeq
where $T_a = \lambda_a / 2$.  In light of our prior discussion, we see that the equilibrium value of $\mu_a$ (i.e., the value required to achieve color neutrality) is proportional to $\braket{A^0_a}$ in QCD.  In both the 2SC and ACFL phases red and green quarks pair symmetrically, so we need only include $\mu_8$, in order to ensure that $n_8 = n_r + n_g - 2 n_b = 2 (n_r - n_b) = 0$.  Thus, we modify the Lagrangian from Eq. (\ref{eq:Lagrangian}) to
\beq
\Lc & = & \overline{q} (i \slashed{D} - \hat{m} + \mu \gamma^0 + \mu_8 \lambda_8 \gamma^0) q + \Lc^{(4)} + \Lc^{(6)}   \nonumber   \\
    && \hspace{5mm} - \Uc (\Phi,\overline{\Phi},T)   .
\eeq

In order to obtain the locally color neutral phase diagram we now minimize the thermodynamic potential with respect to the condensates $\sigma_i$ and $d_i$, and the Polyakov loop variable $\phi_3$ as before, while imposing the additional neutrality constraint
\beq
n_8 = - \frac{\partial \Omega}{\partial \mu_8} = 0   ,
\eeq
as well as the stability condition
\beq
\frac{\partial n_8}{\partial \mu_8} = - \frac{\partial^2 \Omega}{\partial \mu^2_8} > 0   .
\eeq
Thus, our solution is a saddle point of $\Omega$, minimized with respect to the condensates, and maximized with respect to $\mu_8$.

Due to the computational intensity of the saddle point problem for our eight-variable thermodynamic potential, we defer a complete assessment of the effects of color neutrality, together with the strange quark mass and confinement, to a future publication.  However, we report three important results from the \textit{massless} quark limit at $T = 0$, which give insight into the structure of the full color neutral QCD phase diagram.  Note that, as shown in Fig.~\ref{fig:PNJLdvsT}, at $T = 0$ the quark pairing asymmetry vanishes so that a true CFL phase is obtained.

First, in the massless quark limit the color neutrality constraint eliminates the 2SC phase from a large portion of the phase diagram, in favor of the ACFL phase.  We can understand this effect by considering Fig.~\ref{fig:omegas}.  At $\mu = 275$ MeV the thermodynamic potentials of the color neutral NG and CFL phases ($\mu_8 = 0$) are nearly equal, indicating the location of a phase transition between the NG phase, which exists at low $\mu$, and the CFL phase, which exists at high $\mu$.  As the system moves to higher densities the energy of the NG phase is essentially constant, while both the 2SC and CFL phases decrease in energy, becoming more favorable.

The crucial effect of the color neutrality constraint is visible in the thermodynamic potentials at $\mu = 285$ MeV.  In the absence of a color neutrality constraint ($\mu_8 = 0$), we find that the 2SC phase is indeed the lowest energy, and therefore the preferred, phase of the system.  However, in imposing color neutrality, we require the 2SC phase to take on a nonzero $\mu_8 \approx - 40$ MeV, which results in a (physical) 2SC state which is formally higher in energy than the colored state.  This ``additional" energy is sufficient to raise $\Omega_{2SC}$ above both $\Omega_{NG}$ and $\Omega_{CFL}$, with the lower energy CFL phase being the color neutral ground state.  As the system moves to yet higher $\mu$, both the 2SC and CFL phases continue to move to lower energies, with the latter always maintaining a slight energetic advantage.  Thus, at $T = 0$ color neutrality eliminates the 2SC phase altogether.  We do note, however, that the 2SC phase is not eliminated altogether, and that its color neutral form (with $\mu_8 \neq 0$) remains the preferred phase in some portions of the phase diagram for $T > 0$.

A second effect of the local color neutrality constraint is the elimination of the ACFL phase at high $\mu$, in favor of a symmetric CFL phase.  This is somewhat encouraging, given our expectation of a CFL phase at asymptotically high $\mu$, due to general considerations~\cite{Srednicki,Alford1999}.  Thus, we find that the color neutrality constraint disfavors both the asymmetric 2SC and ACFL phases and, at least for some parameters, leads to the complete exclusion of these phases.

A third important effect of color neutrality, which is a corollary of the suppression of the 2SC and ACFL phases, is the ``reemergence" of a low temperature critical point~\cite{Powell,Hatsuda1}.  As shown in~\cite{Basler}, when one allows for 2SC quark pairing (rather than simply a CFL structure) in the absence of a local color neutrality constraint, this critical point is eliminated in favor of a second-order NG-2SC phase transition at intermediate $\mu$.  However, with the 2SC and ACFL phases eliminated at $T = 0$ by the color neutrality constraint, the system once again realizes quark-hadron continuity via a smooth crossover between the NG and CFL phases at low temperatures.

In the case of realistic quark masses, the 2SC phase remains intact after imposing local color neutrality, but the location of the low temperature NG-2SC transition is moved to the right by $\Delta \mu \approx 30$ MeV.  This shift in the phase boundary is not surprising in light of the additional energy required to maintain a nonzero $\mu_8$.  Indeed, the 2SC phase still becomes more favorable (i.e., $\Omega_{2SC}$ decreases) as the system moves to higher $\mu$, but the nonzero value of $\mu_8$ results in an overall shift of $\Omega_{2SC}$ to larger values.  As a result, the NG-2SC phase boundary defined by $\Omega_{2SC} = \Omega_{NG}$ is shifted to larger $\mu$.

\section{Summary}

\begin{figure}
\includegraphics[width=0.45\textwidth]{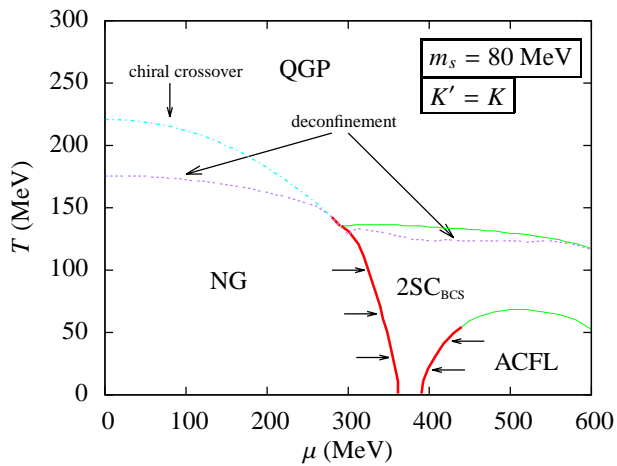}
\caption{(color online).  Proposed phase diagram for three-flavor QCD with spatially homogeneous color neutral phases.  The arrows indicate the movement of the phase boundaries due to the enforcement of local color neutrality [compare to Fig.~\ref{fig:Kp1}(d)].}.
\label{fig:RealPDGuess}
\end{figure}

We have investigated the effects of confinement and realistic mass quarks on the QCD phase diagram, particularly the preferred quark pairing structures at intermediate chemical potentials.  While many prior studies have not enforced local color neutrality in spatially homogeneous phases of asymmetric quark pairing (e.g., 2SC) they have nonetheless assumed a CFL pairing structure at large $\mu$.  Rather, we have shown that in the absence of a local color neutrality constraint the Polyakov loop can give rise to an ACFL phase in which all quark flavors pair, but with unequal magnitudes.  This ACFL phase, which can exist even for three equal mass quarks, provides a mechanism for moving continuously from a true CFL phase to a 2SC phase, as two of the pairing amplitudes (namely, those involving strange quarks) vanish via second-order phase transitions with increasing temperature.

When local color neutrality is enforced, we have shown that the 2SC phase is partially suppressed due to the energy cost of forming the gluon condensate (or $\mu_8 = g \braket{A^0_8}$) required to achieve color neutrality.  However, the ACFL phase remains intact and does not require a gluon condensate (i.e., $\mu_8 = 0$), as the combined effects of the unequal masses and pairing amplitudes dynamically achieve color neutrality.  The mechanism for achieving spatially uniform pairing between quark flavors with imbalanced Fermi seas is the breached pairing described in~\cite{Gubankova2003,Liu2004,Forbes2005}.

While an exhaustive analysis of the effects of color neutrality and realistic quark masses is beyond the scope of the present paper, based on the results obtained here we can propose an educated hypothesis for the QCD phase diagram under the assumptions adopted here, namely, the restriction to spatially homogeneous phases.  Figure~\ref{fig:RealPDGuess} shows our proposed phase diagram, which should be compared to Fig.~\ref{fig:Kp1}(d), in which color neutrality was not enforced.  We expect the essential effect of the color neutrality constraint to be the reduction in size of the 2SC phase, due to the additional energy required to generate the neutralizing gluon condensate.  In particular, the lines of first-order NG-2SC and ACFL-2SC transitions will encroach upon the 2SC region due to the upward shift of $\Omega_{2SC}$, as indicated by the arrows in Fig.~\ref{fig:RealPDGuess}.  Second-order transition lines, however, will remain largely unaffected as $\mu_8 \to 0$ on these boundaries.

A number of outstanding questions exist regarding the PNJL model and the QCD phase diagram which we will address in a future publication~\cite{futurepub}.  Foremost among them is a complete construction of the QCD phase diagram which incorporates local color neutrality along with realistic quark masses.  Also, the effects of charge neutrality and $\beta$ equilibrium, which are important in the study of stable quark matter at low temperatures in neutron stars, remain to be completely elucidated in the context of the PNJL model.

\section{Acknowledgements}
This research was supported in part by NSF Grant No. PHY09-69790.  The authors thank Professor Tetsuo Hatsuda for his insightful comments and discussion.  P.P. thanks Tomoki Ozawa for a helpful discussion regarding symmetry breaking patterns.


\begin{thebibliography}{99}
\bibitem{Enlarged}{Note that in our prior work~\cite{Powell}, we did not observe such an increase in the region of chiral symmetry breaking.  This was due to our defining the deconfinement transition as the (slight) discontinuity in $d\Phi/dT$, which coincides with the chiral phase transition, rather than its maximum finite value.  Here, we use the latter definition as it more naturally extends to the case of non-zero quark masses in which $d\Phi/dT$ is continuous and the deconfinement ``transition" is unquestionably a smooth crossover.}
\bibitem{symmetryfoot}While one often refers to a CFL phase in the case of unequal strange quark mass, it is implicitly assumed that $\mu \gg m_s$ so that $d_1 = d_2 \approx d_3$.  If $\mu$ is not much larger than $m_s$ then the quark pairing amplitudes will no longer be approximately equal and the CFL phase gives way to the ACFL phase.
\end{thebibliography}
\end{document}